\documentclass[12pt]{article}
\usepackage{amsmath, amssymb, graphics, setspace}
\usepackage{amsmath}
\usepackage{amsxtra}
\usepackage{amstext}
\usepackage{amssymb}
\usepackage{latexsym}
\usepackage{dsfont} 
\usepackage{graphicx}

\newcommand{\be}{\begin{equation}}
\newcommand{\ee}{\end{equation}}
\newcommand{\bea}{\begin{eqnarray}}
\newcommand{\eea}{\end{eqnarray}}

\begin{document}

\begin{titlepage}
\begin{center}

\hfill SISSA 36/2011/FM-EP\\

\vskip .3in \noindent


{\Large \bf{The Liouville side of the Vortex}} \\

\vskip .2in

{Giulio Bonelli$^{\heartsuit}$, Alessandro Tanzini and Jian Zhao}

\vskip .05in
{\em\small
International School of Advanced Studies (SISSA) \\via Bonomea 265, 34136 Trieste, Italy \\
and \\ INFN, Sezione di Trieste }
\vskip .05in
{\em\small and}
\vskip .05in
{\em\small
$^{\heartsuit}$
I.C.T.P. \\ Strada Costiera 11, 34014 Trieste, Italy }

\vskip .5in
\end{center}
\begin{center} {\bf ABSTRACT }
\end{center}
\begin{quotation}\noindent

We analyze conformal blocks with multiple (semi-)degenerate field insertions in Liouville/Toda conformal field
theories an show that their vector space is fully reproduced by the four-dimensional limit of open topological string amplitudes on the strip
with generic boundary conditions associated to a suitable quiver gauge theory.
As a byproduct we identify the non-abelian vortex partition function with a specific fusion channel of degenerate conformal blocks.

\end{quotation}
\vfill
\eject

\end{titlepage}

\tableofcontents
\section{Introduction}

The connection between
refined BPS counting in four dimensional ${\cal N}=2$ quiver gauge theories -- namely Nekrasov partition functions \cite{Nek} --
and
Logarithmic Conformal Field Theories on Riemann Surfaces \cite{BPZ},
which was originally noticed in \cite{AGT},
opened a renewed perspective on both the areas.
This correspondence can be studied via geometric engineering \cite{vafa} where
topological strings \cite{topos} can be used to exactly describe the
BPS protected sectors of the gauge theory both at perturbative and non perturbative level
and realizes the above program in M-theory \cite{four,N=2}.

In this context, in \cite{bucov,BTZ} the role of vortex counting was noticed and
proposed to encode in a two dimensional field theoretic perspective the insertion of surface operators
of the type discussed in \cite{surface,KPW,also}.
The role of non-abelian vortices was exploded in \cite{BTZ} by relating their partition function
\cite{shadchin}
with instanton counting and topological string amplitudes.

The Liouville/Toda descriptions of some of these amplitudes with suitable boundary conditions
were provided in terms of insertions of multiple degenerate fields.
In presence of more than one insertion, the conformal blocks span a vector space whose dimension is fixed
by the fusion rules of a generic primary with degenerate fields.
One of the main results of this paper is to provide a full realization of the above in terms of
topological string amplitudes with general boundary conditions.
As a byproduct we identify the non-abelian vortex partition function with a specific fusion channel of the degenerate conformal block,
different to the one considered so far in the literature.

Moreover, we realize the vortex counting problem
as sub-counting instantons by showing how to relate the Nekrasov partition function
and its vortex counterpart by a particular choice of mass parameters in an appropriately
engineered gauge theory in four dimensions. On the gauge theory side, this boils down to consider
surface operator insertions in a theory with a simpler quiver structure.
On the AGT dual side, we notice that the above mass parameters assignments
produces the insertion of degenerate fields in the Liouville/Toda CFT amplitudes.
We study this correspondence in depth, reproduce some known results
and embed them in a wider framework. In particular we show the correspondence between the fusion channel
choice in the Liouville/Toda field theory side and the choice of possible surface operator insertions.
The relation with topological strings, in the form of related strip amplitudes \cite{tv,Iqbal}, is also considered
in full generality for the $SU(2)$ case and in some particular exemplificative ones for $SU(N)$.

We organize our paper as follows.
In section 2 we calculate the CFT dual of $SU(2)$ vortex partition functions.
In section 3 we extend the CFT dual for $SU(N)$ vortices and argue its validity for general strip amplitudes in section 4.
Section 5 contains our conclusion, while some technical details are left for the Appendix.

\section{SU(2) Vortices And Degenerate States}

\subsection{General Setup}

\begin{figure}
\begin{center}
\includegraphics[ width=0.8\textwidth]{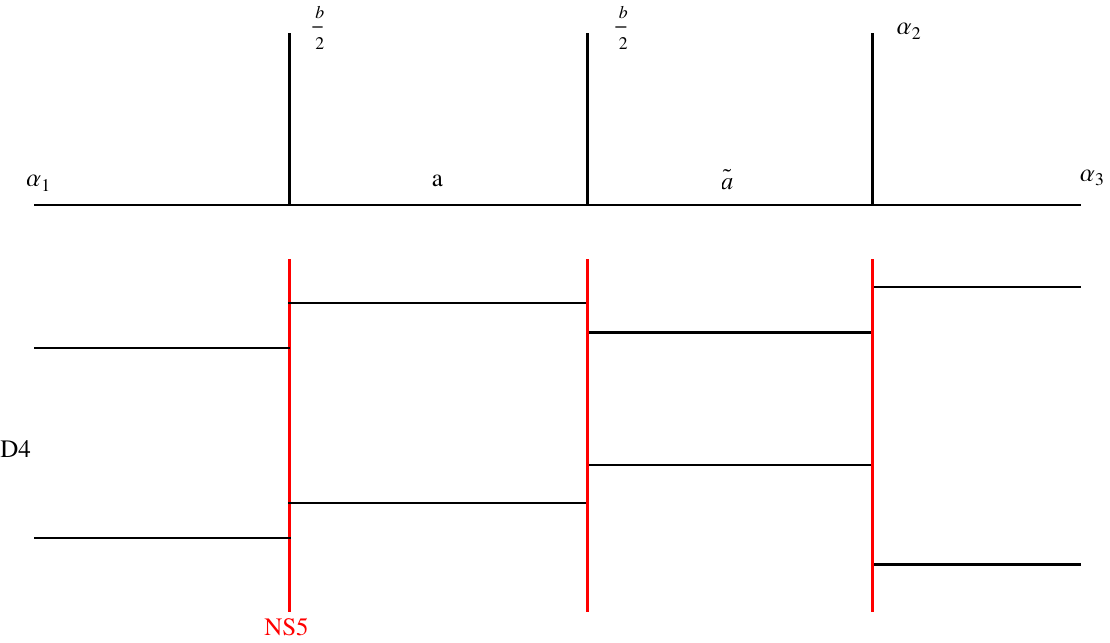}
\caption{AGT relation between $SU(2)$ quiver gauge theory and CFT}
\end{center}
\label{fig:su2brane}
\end{figure}

We start from two node \(SU(2)\) theory with specific parameters.
Its Liouville conformal block dual and brane construction is illustrated in Fig.1.
Following the results of our previous paper \cite{BTZ}, we will focus on the free field limit,
$\epsilon _+\text{:=}\epsilon _1+\epsilon _2=0$.
The parameters of this two node quiver are: $\mu_1,\mu_2$ are masses of antifundamental hypermultiples;
$\mu_3,\mu_4$ are masses of fundamental hypermultiples; $m$ is the mass of bifundamental hypermultiplet and $(a_1,a_2)=(a,-a) ; (\tilde{a}_1,\tilde{a}_2)=(\tilde{a},-\tilde{a})$ are Coulomb branch parameters of the first and second gauge factor.
On the conformal field theory side $\alpha_1,\alpha_2,\alpha_3$ are the external momenta in Liouville theory.
When all parameters are generic, what we get is just the standard AGT correspondence between instanton partition functions of
quiver gauge theories and conformal blocks with five operator insertions.
When there are degenerate states, different fusion channels will give different results which also have different gauge theory interpretation
as we will show in the following.
For two node $SU(2)$ quiver theories there are two channels,
one corresponding to $SU(2)$ vortex partition functions while the other to a simple surface operator as discussed in \cite{KPW}.
The general situation with the insertion of more degenerate fields is discussed in subsequent sections.

The standard AGT-relation \cite{AGT} gives the following map between parameters:
\begin{eqnarray}
\mu _1&=&\alpha _1-\frac{\epsilon _2}{2}  \nonumber\\
\mu _2&=&-\alpha _1-\frac{\epsilon _2}{2}  \nonumber\\
\mu _3&=&\alpha _2+\alpha _3  \\
\mu _4&=&\alpha _2-\alpha _3   \nonumber\\
m&=&-\frac{\epsilon _2}{2}  \nonumber
\end{eqnarray}

The
fusion rules of Liouville field theory imply that
\begin{eqnarray}
\alpha _1=a-s_1\frac{\epsilon _2}{2}  \\
\tilde{a}=a-s_2\frac{\epsilon _2}{2}    \nonumber
\end{eqnarray}
where $s_1$ and $s_2$ are $\pm 1$.
This fixes the masses to
\begin{eqnarray}
\mu _1=a-\left(s_1+1\right)\frac{\epsilon _2}{2} \nonumber\\
\mu _2=-a+\left(s_1-1\right)\frac{\epsilon _2}{2} \nonumber
\end{eqnarray}

Let us remark that
when the differences between Coulomb branch parameters and fundamental/bifundamental masses are linear in $\epsilon_1$ and $\epsilon_2$ the instanton partition function is largely simplified.
To see this let us recall the contribution from antifundamental fields
\be
Z_{\text{antifund}}(m,a,Y)=\prod _{\alpha =1}^2\prod _{(i,j)\in Y_{\alpha }} \left(a_{\alpha }+m+\epsilon _2(j-i)\right)
\ee
Where $a_1=a;a_2=-a$, and $(i,j)$ are the box location in the Young tableaux.
If we choose $s_1=-1$, then $\mu _1=a;\mu _2=-a-\epsilon _2$. The above formula then implies that $Y_2=\emptyset$ and $Y_1$ to be a row.
The other choice $s_1=1$ just exchanges the roles of $Y_1$ and $Y_2$.
So the choice of fusion channel here is just a convention. What is really relevant is the choice of $s_2$.

Let us notice that bifundamental masses can transfer degeneration between adjacent gauge factors of a quiver theory.
Indeed the contribution of bifundamental hypermultiples is:
\bea
&&Z_{\text{bifund}}\left(m,a,\tilde{a},Y,W\right)=\prod _{\alpha =1}^2\prod _{\beta =1}^2Z_{\text{bifund}}^{(\alpha ,\beta )}   \\
&&Z_{\text{bifund}}^{(\alpha ,\beta )}=\prod _{s\in Y_{\alpha }} \prod _{t\in W_{\beta }} \left(m_{\alpha ,\beta }+\epsilon _2\left(A_{Y_{\alpha }}(s)+L_{W_{\beta }}(s)+1\right)\right)\left(m_{\alpha ,\beta }-\epsilon _2\left(A_{W_{\beta }}(t)+L_{Y_{\alpha }}(t)+1\right)\right)  \nonumber\\
&&
m_{\alpha ,\beta }\text{:=}a_{\alpha }-\tilde{a}_{\beta }-m \nonumber
\eea
From the second fusion relation in the diagram one gets
\bea
&&m_{1,1}=\left(s_2+1\right)\frac{\epsilon _2}{2}   \nonumber \\
&&m_{2,2}=\left(1-s_2\right)\frac{\epsilon _2}{2}    \\
&&m_{1,2}=2a+\left(1-s_2\right)\frac{\epsilon _2}{2} \nonumber\\
&&m_{2,1}=-2a+\left(1+s_2\right)\frac{\epsilon _2}{2} \nonumber
\eea

Moreover, AGT-correspondence implies that, up to a $U(1)$ factor which doesn't play any role here,
\bea
&&Z_{\text{Quiver}}\left(a,\tilde{a}=a-s_2\frac{\epsilon _2}{2};\mu _1=a,\mu _2=-a-\epsilon _2;\alpha _2+\alpha _3,\alpha _2-\alpha _3\right)\nonumber\\
&&=\mathcal{F}\left(a+\frac{\epsilon _2}{2},\frac{-\epsilon _2}{2},a,\frac{-\epsilon _2}{2},a-s_2\frac{\epsilon _2}{2},\alpha _2,\alpha _3\right)
\eea
where the LHS is the instanton partition function of $SU(2)$ quiver gauge theory and the RHS is the conformal block of Liouville field theory.

In the following we will show that
when $s_2=-1$
the quiver partition function in the
above formula reduces to the $SU(2)$ vortex partition function, while when $s_2=1$, it corresponds to the
$SU(2)$ simple surface operator.

\subsection{SU(2) Vortices}

Let us start investigating the case \(s_2=-1\) where
\bea
&&m_{1,1}=0  \nonumber\\
&&m_{2,2}=\epsilon _2\nonumber\\
&&m_{1,2}=2a+\epsilon _2\nonumber\\
&&m_{2,1}=-2a
\label{param}
\eea

\begin{figure}
\begin{center}
\includegraphics[ width=0.8\textwidth]{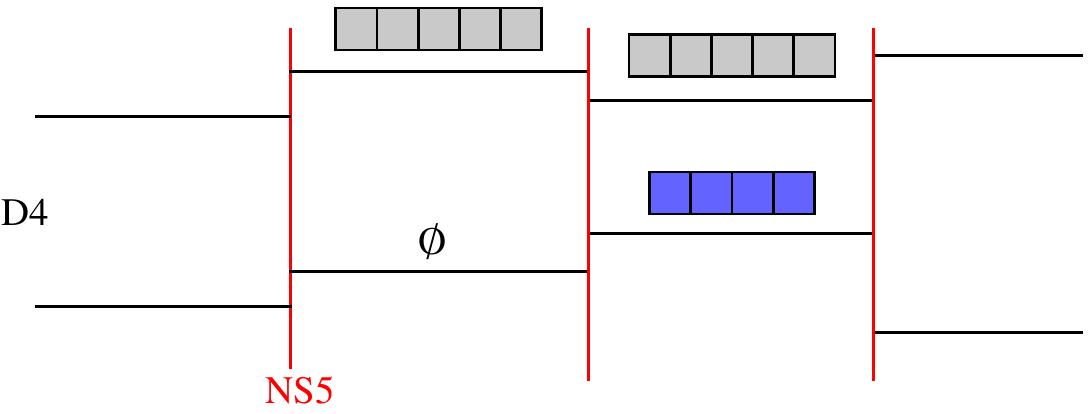}
\caption{ $SU(2)$  vortex  from quiver theory}
\end{center}
\label{fig:su2vortex_brane}
\end{figure}

To start with, let's focus on $Z_{\text{bifund}}^{(1,1)}$ :
\bea
&&Z_{\text{bifund}}^{(1,1)}=\prod _{s\in Y_1} \left(\epsilon _2\left(A_{Y_1}(s)+L_{W_1}(s)+1\right)\right)\prod _{t\in W_1} \left(-\epsilon _2\left(A_{W_1}(t)+L_{Y_1}(t)+1\right)\right)  \nonumber \\
&&\text{where:} \\
&&\prod _{t\in W_1} \left(-\epsilon _2\left(A_{W_1}(t)+L_{Y_1}(t)+1\right)\right)=\prod _{(i,j)\in W_1} -\epsilon _2\left(A_{W_1}(i,j)+L_{Y_1}(i,j)+1\right)\nonumber
\label{vortex}
\eea

From the discussion of the previous section we know that the choice of the fundamental mass parameter in (\ref{param}) implies that
  \(Y_1\) is a row diagram. Moreover, from the results in the Appendix, one gets that the bifundamental masses in (\ref{param}) set
also \(W_1\) to be a row of the same length which we call $k_1$, see the Fig.2.

To simplify the formul\ae, let's define some notations:
\be
(x)_{Y,W}\text{:=}\prod _{s\in Y} \prod _{t\in W} \left(x+\epsilon _2\left(A_{Y}(s)+L_{W}(s)+1\right)\right)\left(x-\epsilon _2\left(A_{W}(t)+L_{Y}(t)+1\right)\right)
\ee
and
\bea
(x)_Y&\text{:=}&(x)_{Y,\emptyset } \nonumber\\
H_Y&\text{:=}&(0)_{Y,\emptyset } \nonumber\\
(x)_k&\text{:=}&(x)_{\emptyset ,\left(1^k\right)} \nonumber\\
(x)_{k_1,k_2}&\text{:=}&(x)_{\left(1^{k_1}\right),\left(1^{k_2}\right)} \nonumber
\eea

Let's calculate  \(Z_{\text{bifund}}^{(1,1)}\) explicitly
\be
Z_{\text{bifund}}^{(1,1)}=\prod _{i=1}^{k_1}\epsilon _2 i \prod _{j=1}^{k_1}-\epsilon _2j=\left(\epsilon _2\right)_{k_1}^2(-1)^{k_1}
\ee
The contribution form $Z_{\text{bifund}}^{(2,2)}$ is instead
\be
Z_{\text{bifund}}^{(2,2)}=\prod _{t\in W_2} -\epsilon _2\left(A_{W_2}(t)+L_{\emptyset }(t)\right)=\prod _{(i,j)\in W_2} -\epsilon _2(j-i-1)
\ee
which is non zero only if \(W_2\) is a row. Let's denote its length by \(k_2\).
Then
\be
Z_{\text{bifund}}^{(2,2)}=\left(\epsilon _2\right)_{k_2}
\ee

By including the contributions from \(Z_{\text{bifund}}^{(1,2)}\) and  \(Z_{\text{bifund}}^{(2,1)}\)
we get the final formula
\be
Z_{\text{bifund}}=\left(\epsilon _2\right)_{k_1}^2(-1)^{k_1}\left(\epsilon _2\right)_{k_2}(-2a)_{k_1}\left(2a+\epsilon _2\right)_{k_1,k_2}
\ee

The contribution from the anti-fundamental matter can be computed with analogous methods giving
\be
Z_{\text{antifund}}=\left(\epsilon
_2\right)_{k_1}(-2a)_{k_1}
\ee

The generic
contribution from the vector multiplets is
\bea
&&Z_{\text{vect}}(a,Y)=\prod _{\alpha ,\beta =1}^2Z_{\text{vect}}^{(\alpha ,\beta )}(a,Y)\\
&&Z_{\text{vect}}^{(\alpha ,\beta )}(a,Y)=\prod _{s\in Y_{\alpha }} \prod _{t\in Y_{\beta }} \left(a_{a,\beta }+\epsilon _2\left(A_{Y_{\alpha }}(s)+L_{Y_{\beta
}}(s)+1\right)\right)^{-1}\left(a_{\alpha ,\beta }-\epsilon _2\left(A_{Y_{\beta }}(t)+L_{Y_{\alpha }}(t)+1\right)\right)^{-1}   \nonumber \\
&&a_{\alpha ,\beta }\text{:=}a_{\alpha }-a_{\beta } \nonumber
\eea
which reduces for the first node of our specific Young tableaux to
\be
Z_{\text{vect}}(a,Y)=\left(\left(\epsilon _2\right)_{k_1}(-2a)_{k_1}\right){}^{-2}
\ee

The fundamental matter
\(Z_{\text{fund}}\) is in the standard form
\be
Z_{\text{fund}}=\left(\tilde{a}+\mu _3\right)_{k_1}\left(\tilde{a}+\mu _4\right)_{k_2}\left(-\tilde{a}+\mu _3\right)_{k_1}\left(-\tilde{a}+\mu _4\right)_{k_2}
\ee
while the contribution from the second gauge factor of the quiver is:
\be
Z_{\text{vect}}(\tilde{a},W)=\frac{(-1)^{k_1+k_2}}{\left(\epsilon _2\right)_{k_1}^2\left(\epsilon _2\right)_{k_2}^2\left(2a+\epsilon _2\right)_{k_1,k_2}\left(-2\tilde{a}\right)_{k_2,k_1}}
\ee


In summary, the total partition function of the quiver theory with specific choice of masses reads
\be
Z_{\text{Quiver}}(k_1,k_2)
=\frac{(-1)^{k_1}\left(\tilde{a}+\mu _3\right)_{k_1}\left(\tilde{a}+\mu _4\right)_{k_2}\left(-\tilde{a}+\mu _3\right)_{k_1}\left(-\tilde{a}+\mu _4\right)_{k_2}}{\left(\epsilon _2\right)_{k_1}\left(\epsilon
_2\right)_{k_2}\left(2\tilde{a}\right)_{k_1,k_2}}
\ee

This, up-to a sign factor which can be absorbed in the vortex counting parameter coincides to\footnote{With respect to \cite{BTZ} we set
$\hbar=-\epsilon_2$. These sign factors will be disregarded in the following without further notice.}
the $SU(2)$ vortex partition function studied in \cite{BTZ}:
\be
Z_{\text{vortex}}^{\text{SU}(2)}(\pmb{k})=\frac{(-1)^{k_2}\left(a-m_1\right)_{k_1}\left(-a-m_1\right)_{k_2}\left(a-m_2\right)_{k_1}\left(-a-m_2\right)_{k_2}}{\left(\epsilon _2\right)_{k_1}\left(\epsilon _2\right)_{k_2}\left(a_{1,2}\right)_{k_1,k_1}}
\ee
Notice that we should identify $m_i=-\mu_{i+2}$,and $\tilde{a}$ as $a$, since it is the second gauge factor that couples to hypermultiplets with generic masses.

To conclude the matching, notice that in the
two nodes quiver theory, we have two parameters \(q_1,q_2\) which are the exponential of the gauge couplings of the quiver theory.
These are related to the
vortex counting parameters \(z_1,z_2\) of vortex partition functions as
\be
q_1^{k_1}\left(q_2\right){}^{k_1+k_2}=\left(q_1q_2\right){}^{k_1}q_2^{k_2}=z_1^{k_1}z_2^{k_2}
\ee
From the CFT viewpoint $z_i$ are the insertion points of the degenerate fields.

\subsection{SU(2) simple surface operators}
A natural question is to find what's the result in the other channel.
As expected we find it is the result of \cite{KPW}. So, let's now choose \(s_2=1\), then
\bea
\tilde{a}&=&a-\frac{\epsilon _2}{2}\nonumber\\
m_{1,1}&=&\epsilon _2\nonumber\\
m_{2,2}&=&0\nonumber\\
m_{1,2}&=&2a  \nonumber\\
m_{2,1}&=&-2a+\epsilon _2
\eea

\begin{figure}
\begin{center}
\includegraphics[ width=0.8\textwidth]{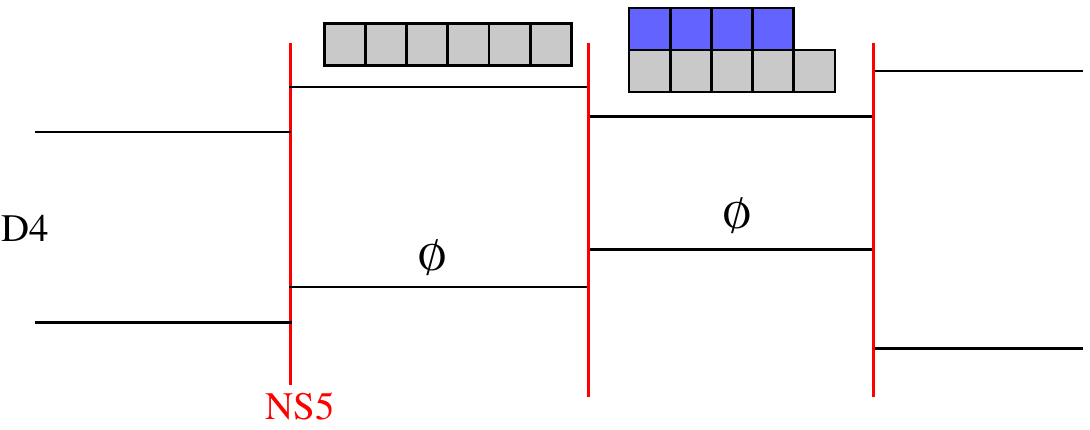}
\caption{ $SU(2)$  simple surface operators from quiver theory}
\end{center}
\label{fig:su2surface_brane}
\end{figure}

In this case, the contribution of the bifundamentals reads
\bea
&&Z_{\text{bifund}}^{(1,1)}=\prod _{s\in Y_1} \left(\epsilon _2\left(A_{Y_1}(s)+L_{W_1}(s)+2\right)\right)\prod _{t\in W_1} \left(-\epsilon _2\left(A_{W_1}(t)+L_{Y_1}(t)\right)\right)\\
&& \text{where:}\\
&&\prod _{t\in W_1} \left(-\epsilon _2\left(A_{W_1}(t)+L_{Y_1}(t)\right)\right)=\prod _{(i,j)\in W_1} -\epsilon _2\left(A_{W_1}(i,j)+L_{Y_1}(i,j)\right) \nonumber
\label{surface}
\eea

Using once again the results in the appendix, the bifundamental contribution
%
%
%
%
\be
Z_{\text{bifund}}^{(2,2)}=\prod _{t\in W_2} \left(-\epsilon _2\left(A_{W_2}(t)+L_{\emptyset }(t)+1\right)\right)=\prod _{t\in W_2} \left(-\epsilon
_2(j-i)\right)
\ee
is non vanishing only if \(W_2=\emptyset\), see Fig.3.

Therefore, the bifundamental contributions
are given by
\bea
&&Z_{\text{bifund}}^{(1,1)}=H_{Y_1}H_{W_1}(-1){}^{k_1}  \nonumber\\
&&Z_{\text{bifund}}^{(2,2)}=1\nonumber\\
&&Z_{\text{bifund}}^{(1,2)}=(-1)^{k_1+1}(-2a)_{k_1+1}\nonumber\\
&&Z_{\text{bifund}}^{(2,1)}=(-1)^{k_1+k_2}\left(\tilde{a}_{1,2}\right)_{W_1}\nonumber
\eea

The contribution from the other factors can be analogously derived to be
\be
Z_{\text{antifund}}Z_{\text{vect}}(a,Y)=\frac{1}{\left(\epsilon _2\right)_{k_1+1}(-2a)_{k_1+1}}
\ee
\be
Z_{\text{vect}}\left(\tilde{a},W\right)=\frac{1}{\left(H_{W_1} \left(\tilde{a}_{1,2}\right)_{W_1}\right){}^2}
\ee
\be
Z_{\text{fund}}=\left(\tilde{a}-\mu _3\right)_{W_1}\left(\tilde{a}-\mu _4\right)_{W_1}
\ee
and finally we get
\be
Z_{\text{Quiver}}(W_1)=\frac{(-1)^{k_1+k_2+1}\left(\tilde{a}-\mu _3\right)_{W_1}\left(\tilde{a}-\mu _4\right)_{W_1}}{H_{W_1} \left(\tilde{a}_{1,2}\right)_{W_1}}
\ee
which is
the partition function of $SU(2)$ simple surface operator \cite{KPW}
\be
Z_{\text{simple surface}}=\frac{\left(a+m_1\right)_{W_1}\left(a+m _2\right)_{W_1}}{H_{W_1} \left(a_{1,2}\right)_{W_1}}
\ee
Now the identification of parameters goes as
\be
q_1^{k_1+1}\left(q_2\right){}^{k_1+k_2}=\frac{z_1}{z_2}z_1^{k_1}z_2^{k_2}
\ee
As already noticed, $z_i$ are the insertion points of the degenerate fields.

\subsection{Relation To Open Topological String amplitudes}

The amplitudes discussed in the previous sections can be derived as four dimensional limits of Open Topological String amplitudes on the strip
with suitable boundary conditions
\cite{KPW,BTZ}.
\begin{figure}
\begin{center}
\includegraphics[ width=.5\textwidth]{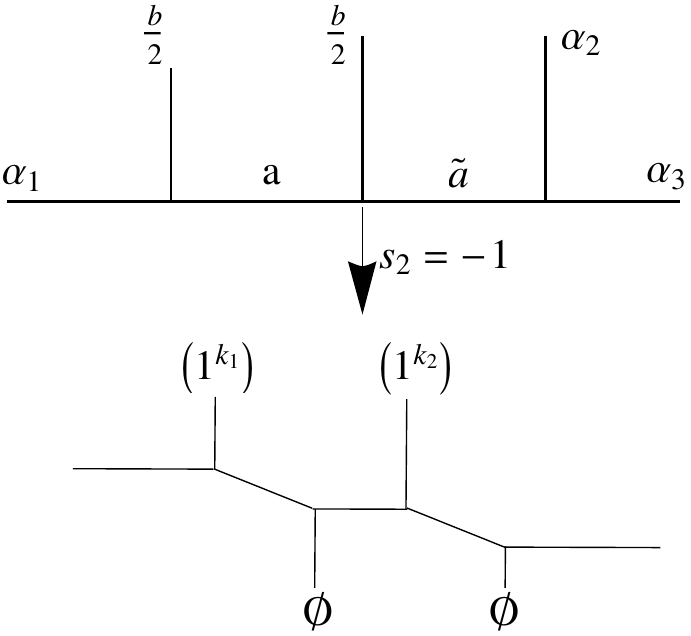}
\caption{relation between $SU(2)$  vortex  and CFT}
\end{center}
\label{fig:su2vortex_strip}
\end{figure}
\begin{figure}
\begin{center}
\includegraphics[ width=.5\textwidth]{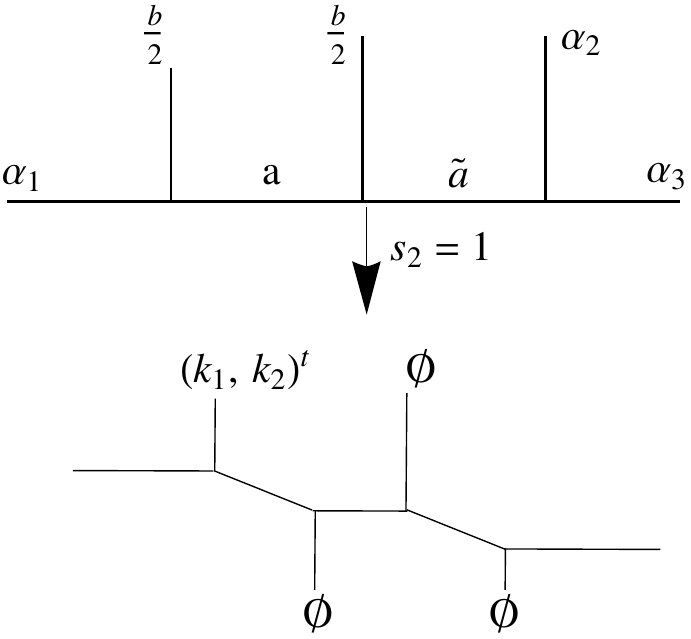}
\caption{relation between $SU(2)$  simple surface operator  and CFT}
\end{center}
\label{fig:su2surface_strip}
\end{figure}
The discussion of the previous section then provides the CFT interpretation of this class of strip amplitudes, as summarized in
Fig.4 and Fig.5.
Actually, this is the simplest  situation.
For example we can have more than two degenerate states, then does this story still holds? The answer is yes.
From our previous calculations, we can deduce three general laws:
(1)the number of nodes of the quiver equals the number of degenerate states.
(2)the total number of rows of Young-tableaux increase by one when counting from left to right along the quiver of gauge theory nodes.
(3)  different fusion channels just tell us on which gauge factor of the quiver to associate an extra row in the partition.
So if we have $n$ degenerate states, the corresponding quiver has $n$ nodes, and on each node there are two choices to add a new row.
For convenience let's define a fusion vector $\mathfrak{V}\in \mathbb{Z}_2^n$,
whose $i$-th component is $1$ if we add a new row onto the partition attached to the first $D4$ brane and $2$ if to the second.
For example, the non-abelian vortex partition function is associated to $\mathfrak{V}=(1,2)$,
while the simple surface operator partition function is associated to $\mathfrak{V}=(1,1)$.

When we have $n$ degenerate states, the Young-tableaux on the final node are a couple $(Y,W)$ satisfying the constraint $n_1+n_2=n$,
where $n_1,n_2$ are respectively the number of rows of $Y$ and of $W$.
Hence we conclude that the four dimensional limit of
the strip amplitudes of the form $A_{\{\emptyset ,\emptyset \}}^{\{Y,W\}}$, that is with boundary conditions labeled by $Y$ and $W$,
reproduces the full conformal block vector space including all the possible fusion channels.
For example we can choose  $\mathfrak{V}=(1,...,1,2,...,2)$, where there are $n_1$ 1's and $n_2$ 2's and can prove explicitly
that for this choice of fusion vector our claim is correct, see Fig.6.
\begin{figure}
\begin{center}
\includegraphics[ width=0.8\textwidth]{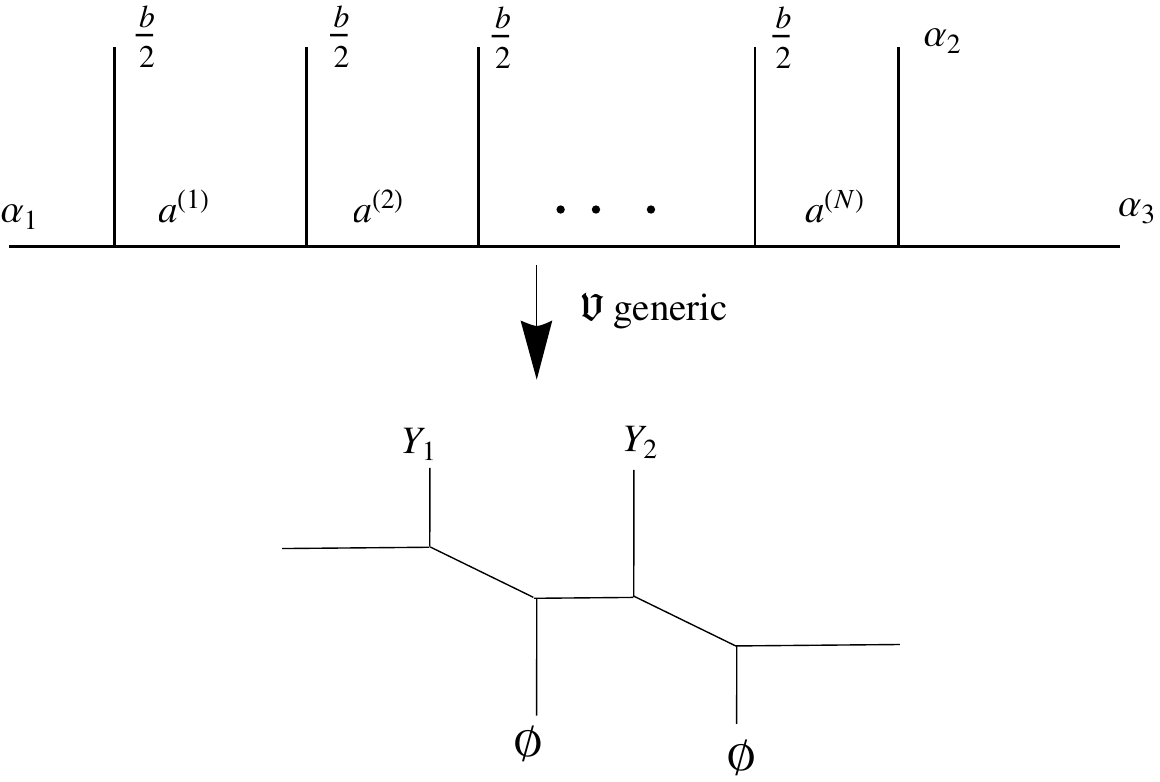}
\caption{relation between $SU(2)$ strip amplitudes and CFT}
\end{center}
\label{fig:su2general}
\end{figure}

\section{SU(N) Generalization}

In the following we will give the natural generalization to $SU(N)$ theories. We know
that the $SU(N)$ vortex partition function should have $N$ independent counting parameters, thus from the previous section's discussion we know
that the associated $SU(N)$ quiver theory will have $N$ nodes.
The quiver configuration reads as the brane construction illustrated in Fig.7.
\begin{figure}
\begin{center}
\includegraphics[ width=.8\textwidth]{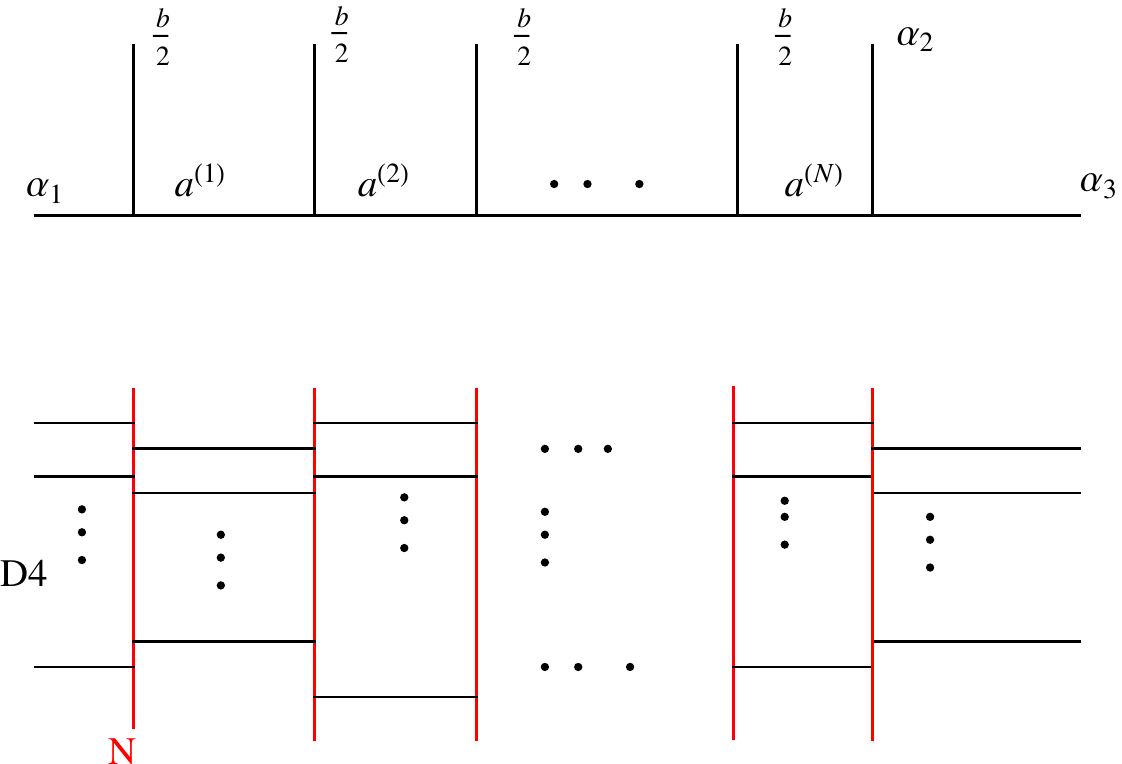}
\caption{AGT relation between $SU(N)$ quiver gauge theory and CFT}
\end{center}
\label{fig:sunbrane}
\end{figure}

\subsection{SU(N) vortices}

The Young-tableaux configuration of quiver gauge theory corresponding to vortex partition function is such that at the
$L$-th node the arrows of Young-tableaux are \(Y^{(L)}=\left(1^{k_1},\text{...},1^{k_L},\emptyset ,\text{...},\emptyset \right)\).
This configuration can be obtained from a given bifundamental mass assignments as displayed in the following.
We will see that this choice of masses correctly reproduces the fusion rules for Toda field theory.

Let us consider the $L$-th node of the quiver and calculate \(Z_L Z_{L,L+1}\), where $Z_L$ is the vector contribution of the $L$-th
node while $Z_{L,L+1}$ the corresponding bifundamental.
Following the arguments in the Appendix, we can read out the $L$-th bifundamental mass to be
\be
m_{\alpha ,\alpha }^{(L)}\text{:=}a_{\alpha }^{(L)}-a_{\alpha }^{(L+1)}-m_L=\delta _{\alpha ,L+1}\epsilon _2
\label{sun bifundmass}
\ee
Then the matrix of masses is given by
\bea
m_{\alpha ,\beta }^{(L)}=\left\{
\begin{array}{cc}
 a_{\alpha ,\beta }^{(L)}=a_{\alpha ,\beta }^{(L+1)} & \alpha \in [1,L];\beta =[1,L] \nonumber\\
 a_{\alpha ,\beta }^{(L+1)} & \alpha \in [1,L];\beta \in [L+1,N] \\
 a_{\alpha ,\beta }^{(L)} & \alpha \in [L+1,N];\beta =[1,L] \nonumber
\end{array}
\right.
\eea
We find it better to write $Z_L$ in three parts according to above mass matrix formula:
\be
Z_L^{-1}= \prod _{\alpha ,\beta =1}^L\left(a_{\alpha ,\beta }^{(L)}\right){}_{k_{\alpha },k_{\beta }}
\prod _{\alpha =1}^L\prod _{\beta =L+1}^N(-1)^{k_{\alpha }}\left(a_{\beta ,\alpha }^{(L)}\right){}_{k_{\alpha }}
\prod _{\beta =1}^L\prod _{\alpha =L+1}^N\left(a_{\alpha ,\beta }^{(L)}\right){}_{k_{\beta }}\nonumber
\ee
Correspondingly, $Z_{L,L+1}$ read
\bea
Z_{L,L+1}&=&
\left\{\prod _{\alpha =1}^L\prod _{\beta =1}^L\left(m_{\alpha ,\beta }^{(L)}\right){}_{k_{\alpha },k_{\beta }}\right\}\times
\nonumber\\
&&\left\{\prod _{\alpha =1}^L\prod _{\beta =L+2}^N(-1)^{k_{\alpha }}\left(-m_{\alpha ,\beta }^{(L)}\right){}_{k_{\alpha }}\right\}\left\{\prod
_{\alpha =1}^L\left(m_{\alpha ,L+1}^{(L)}\right){}_{k_{\alpha },k_{L+1}}\right\}\times \nonumber\\
&& \left\{\prod _{\alpha =L+1}^N\prod _{\beta =1}^L\left(m_{\alpha ,\beta }^{(L)}\right){}_{k_{\beta }}\right\}\left\{\prod _{\alpha =L+1}^N\left(m_{\alpha
,L+1}^{(L)}\right){}_{k_{L+1}}\right\}\nonumber
\eea
Then we get:
\be
Z_L Z_{L,L+1}=\frac{\left\{\prod _{\alpha =1}^L\left(a_{\alpha ,L+1}^{(L+1)}\right){}_{k_{\alpha },k_{L+1}}\right\}}{\left\{\prod _{\alpha =1}^L(-1)^{k_{\alpha }}\left(a_{L+1,\alpha }^{(L)}\right){}_{k_{\alpha }}\right\}}\left\{\prod _{\alpha =L+2}^N\left(a_{\alpha ,L+1}^{(L+1)}\right){}_{k_{L+1}}\right\}\left(\epsilon _2\right)_{k_{L+1}}
\ee

The mass spectrum of the antifundamental hypermultiplets is assigned as
\be
\left(\mu _1,\mu _2,\text{...},\mu _N\right)=\left(-a_1^{(1)}-\epsilon _2,-a_2^{(1)},\text{...},-a_N^{(1)}\right)
\label{sun antimass}
\ee
and the correspondent contribution to the instanton partition function
is
\be
Z_{\text{antifund}}=\prod _{f=1}^N\prod _{i=1}^{k_1}\left(a_1^{(1)}+\mu_f+\epsilon _2(1-i)\right)=(-1)^{N k_1}\left(\epsilon _2\right)_{k_1}\prod
_{i=2}^N\left(a_{i,1}^{(1)}\right){}_{k_1}
\ee
Finally, the vector contribution of the last $N$-th node is
\be
Z_N^{-1}=\prod _{\alpha =1}^N\left(\epsilon _2\right)_{k_{\alpha }}^2(-1)^{k_{\alpha }}\prod _{\alpha <\beta }^N(-1)^{k_{\alpha }+k_{\beta }}\left(a_{\alpha ,\beta }^{(N)}\right)_{k_{\alpha },k_{\beta }}^2
\ee
Then the instanton partition function of this quiver is:
\be
Z_{\text{Quiver}}=\frac{(-1)^{N k_1+\sum _{\alpha }k_{\alpha }}\text{  }\prod _{\alpha ,f=1}^N(-1)^{k_{\alpha }}\left(-a_{\alpha }^{(N)}+\mu _{f+N}\right){}_{k_{\alpha }}}{\prod _{\alpha =1}^N\left(\epsilon _2\right)_{k_{\alpha }}\prod _{\alpha <\beta }^N(-1)^{k_{\beta }}\left(a_{\alpha ,\beta }^{(N)}\right){}_{k_{\alpha },k_{\beta }}}
\ee

Following the result of \cite{BTZ}, and identifing $\hbar=-\epsilon_2$ , the $SU(N)$ vortex partition function can be written as:
\bea
Z_{\text{vortex}}^{\text{SU}(N)}&=&\sum _{\pmb{k}}Z_{\text{vortex}}^{\text{SU}(N)}(\pmb{k})\prod _{i=1}^Nz_i^{k_i}\\
Z_{\text{vortex}}^{\text{SU}(N)}(\pmb{k})&=&\frac{\prod _{\alpha ,f=1}^N(-1)^{k_{\alpha }}\left(-a_{\alpha }-m_f\right)_{k_{\alpha }}}{\prod _{\alpha =1}^N\left(\epsilon _2\right)_{k_{\alpha }}\prod _{\alpha <\beta }^N(-1)^{k_{\beta }}\left(a_{\alpha ,\beta }\right)_{k_{\alpha },k_{\beta }}} \nonumber
\eea
This can be identified with the quiver instanton partition function by setting $a^{(N)}_{\alpha}=a_{\alpha}$ and $m_f=-\mu_{N+f}$.
The counting parameters $z_i$ are identified as
\bea
&&\prod _{i=1}^Nq_i^{\sum _{j=1}^ik_j}=\prod _{i=1}^Nz_i^{k_i}\\
&&z_i\text{:=}\prod _{j=i}^{N+1-i}q_i\nonumber
\eea

In conclusion,
the instanton partition function of quiver gauge theory with \(Y^{(L)}=\left(1^{k_1},\text{..},1^{k_L},\emptyset ,\text{...},\emptyset \right)\)
with parameters in formula (\ref{sun bifundmass}) and (\ref{sun antimass}) gives the $SU(N)$ vortex partition
function.

\subsection{SU(N) Simple Surface Operators}

From the previous arguments we can argue  that
the four dimensional limit of the strip amplitude
$A_{\{\emptyset ,\text{...},\emptyset\}}^{W,\emptyset ,\text{...},\emptyset}$, with $W=\left(k_1,k_2,\text{...},k_N\right)$,
corresponds to the quiver gauge theory with the following
Young-tableaux assignment
\bea
Y^{(L)}&=&\left(Y_L,\emptyset ,\text{...},\emptyset \right)\\
Y_L^t&=&\left(k_1+(N-L),k_2+(N-L),\text{...},k_L+(N-L)\right)\nonumber
\eea
The corresponding bifundamental masses can be obtained by following the arguments displayed in the appendix to be
\be
m_{\alpha ,\alpha }^{(L)}=a_{\alpha }^{(L)}-a_{\alpha }^{(L+1)}-m_L=\delta _{\alpha ,1}\epsilon _2
\label{surface bifund}
\ee
for the $L$-th node.
The corresponding vector contribution for the $L$-th node is
\bea
Z_L^{-1}&=&(-1)^{\left|Y_L\right|}H_{Y_L}^2\prod _{\beta =2}^N\left(a_{1,\beta }^{(L)}\right){}_{Y_L}\prod _{\alpha =2}^N(-1)^{\left|Y_L\right|}\left(a_{1,\alpha
}^{(L)}\right){}_{Y_L}
\eea
while the bifundamental is
\be
Z_{L,L+1}=\left(\epsilon _2\right)_{Y_L,Y_{L+1}}\prod _{\beta =2}^N\left(a_{1,\beta }^{(L)}\right){}_{Y_L}\prod _{\alpha =2}^N(-1)^{\left|Y_{L+1}\right|}\left(a_{1,\alpha
}^{(L+1)}\right){}_{Y_{L+1}}
\ee
so that
\be
\prod _{L=1}^{N-1}Z_LZ_{L,L+1}=\left\{\prod _{L=1}^{N-1}\frac{ \left(\epsilon _2\right)_{Y_L,Y_{L+1}}}{(-1)^{\left|Y_L\right|}H_{Y_L}^2}\right\}\frac{\prod
_{\alpha =2}^N(-1)^{\left|Y_N\right|}\left(a_{1,\alpha }^{(N)}\right){}_{Y_N} }{\prod _{\alpha =2}^N(-1)^{\left|Y_1\right|}\left(a_{1,\alpha }^{(1)}\right){}_{Y_1}}
\ee
Using the result of the last Appendix, we can rewrite
\be
\left(\epsilon _2\right)_{Y_L,Y_{L+1}}=(-1)^{\left|Y_L\right|+L}H_{Y_L}H_{Y_{L+1}}
\ee
and finally get
\be
\prod _{L=1}^{N-1}Z_L Z_{L,L+1}=(-1)^{\sum _{L=1}^{N-1}L}\frac{H_{Y_N}\prod _{\alpha =2}^N(-1)^{\left|Y_N\right|}\left(a_{1,\alpha }^{(N)}\right){}_{Y_N} }{H_{Y_1}\prod _{\alpha =2}^N(-1)^{\left|Y_1\right|}\left(a_{1,\alpha }^{(1)}\right){}_{Y_1}}
\ee
Notice that, as in $SU(2)$ case, the spectrum of antifundamental hypermultiplets is fixed to be the same both for simple surface operator and nonabelian vortices.
What distinguishes the different cases are the different fusion rules channels.
The corresponding factors are then
\bea
Z_{\text{fund}}&=&\prod _{f=1}^N\left(a_1^{(N)}-\mu _{f+N}\right){}_{Y_N}\\
Z_{\text{antifund}}&=&(-1)^{\left|Y_1\right|}H_{Y_1}\prod _{\alpha =2}^N\left(a_{1,\alpha }^{(1)}\right){}_{Y_1}\\
Z_N^{-1}&=&(-1){}^{N\left|Y_N\right|}H_{Y_N}^2\prod _{\alpha =2}^N\left(a_{1,\alpha }\right)_{Y_N}^2
\eea
which finally give
\be
Z_{\text{Quiver}}=(-1)^{\sum _{L=1}^NL+N\left|Y_1\right|+\left|Y_N\right|}\frac{\prod _{f=1}^N\left(a_1^{(N)}-\mu _{f+N}\right){}_{Y_N}}{H_{Y_N}\prod _{\alpha =2}^N\left(a_{1,\alpha }^{(N)}\right){}_{Y_N}}.
\ee
This, after the identifications $\hbar=-\epsilon_2$, $a^{(N)}_1=a_1$, $m_f=-\mu_{f+N}$ and $\lambda=Y_N$, is the simple surface operator
partition function discussed in \cite{KPW} under the same counting parameters identification that we
used in the last section.




\subsection{Toda Fusion Rules From Quiver Gauge Theory}

In this subsection we show how to
\textit{derive} fusion rules of semidegenerate states of Toda field theory
from our construction.
Let's concentrate on the L-th node of the quiver and recall the diagonal part of the mass assignment
\be
m_{\alpha ,\alpha }^{(L)}\text{:=}a_{\alpha }^{(L)}-a_{\alpha }^{(L+1)}-m_L
\ee
By denoting
\({\tt m}_L=m_L(1,1,\text{...},1)\), being a vector of $N$ entries all equal to $m_L$, we can write the above formula as
\be
{\tt a}^{(L)}-{\tt a}^{(L+1)}={\tt m}^{(L)}-{\tt m}_L
\ee
where ${\tt a}^{(L)}$ denotes the vector of internal momenta at the $L$-th node
and
${\tt m}^{(L)}$ the vector of diagonal entries of the mass matrix at the $L$-node.
Actually, for this assignment of external momenta, Toda fusion rules have $N$ channels.
For the i-th channel  \({\tt m}^{(L)}=\epsilon _2 u_i=\epsilon _2\left(u_1-\sum _{j=1}^{i-1}e_j\right)\). Where \(u_i\)
is the unit vector in the i-th direction in \(\mathbb{R}^N\) and \(e_j\text{:=}u_j-u_{j+1}\) are the
simple roots of the \(\mathfrak{s}\mathfrak{l}_N\) algebra.
Then we have
\be
{\tt a}^{(L)}- {\tt a}^{(L+1)}=\epsilon _2\left(u_1-\sum _{j=1}^{i-1}e_j\right)-{\tt m}_L=
\epsilon _2\left(u_1-\frac{{\tt m}_L}{\epsilon _2}\right)-\epsilon _2\sum _{j=1}^{i-1}e_j
\ee
If we set \({\tt m}_L=\epsilon _2\frac{1}{N}(1,1,\text{...},1)\), then
\be
{\tt a}^{(L)}-{\tt a}^{(L+1)}=\epsilon _2\left(-\omega _1\right)-\epsilon _2\sum _{j=1}^{i-1}e_j
\ee
where \(\omega _1\) is the highest weight of the fundamental representation of \(\mathfrak{s}\mathfrak{l}_N\). The above formula can be recognized
as the fusion rule calculated in \cite{FL1}.

For $SU(N)$ N nodes quiver, we can have $N$ semidegenerate states, for each one of them we have $N$ channels.
We can use a $N$-dimensional vector of integer entries $\mathfrak{V}$ to denote the choice of the fusion channels.
The fusion vector $\mathfrak{V}$ is built as follows: if on the $L$-th node we choose $k$-th channel, namely
$m_{\alpha ,\alpha }^{(L)}=\epsilon _2\delta _{\alpha ,k}$, then the corresponding $L$-th component of $\mathfrak{V}$ is set equal to $k$.
For example for the $SU(N)$ vortex $\mathfrak{V}_{\rm vortex}=(1,2,...,N)$, while for $SU(N)$ simple surface operator,
$\mathfrak{V}_{\rm simple surface}=(1,1,...,1)$.

The relation with the four dimensional limit of strip amplitudes goes as in the $SU(2)$ as depicted in Figs.8,9
and Fig.10.

\begin{figure}
\begin{center}
\includegraphics[ width=0.8\textwidth]{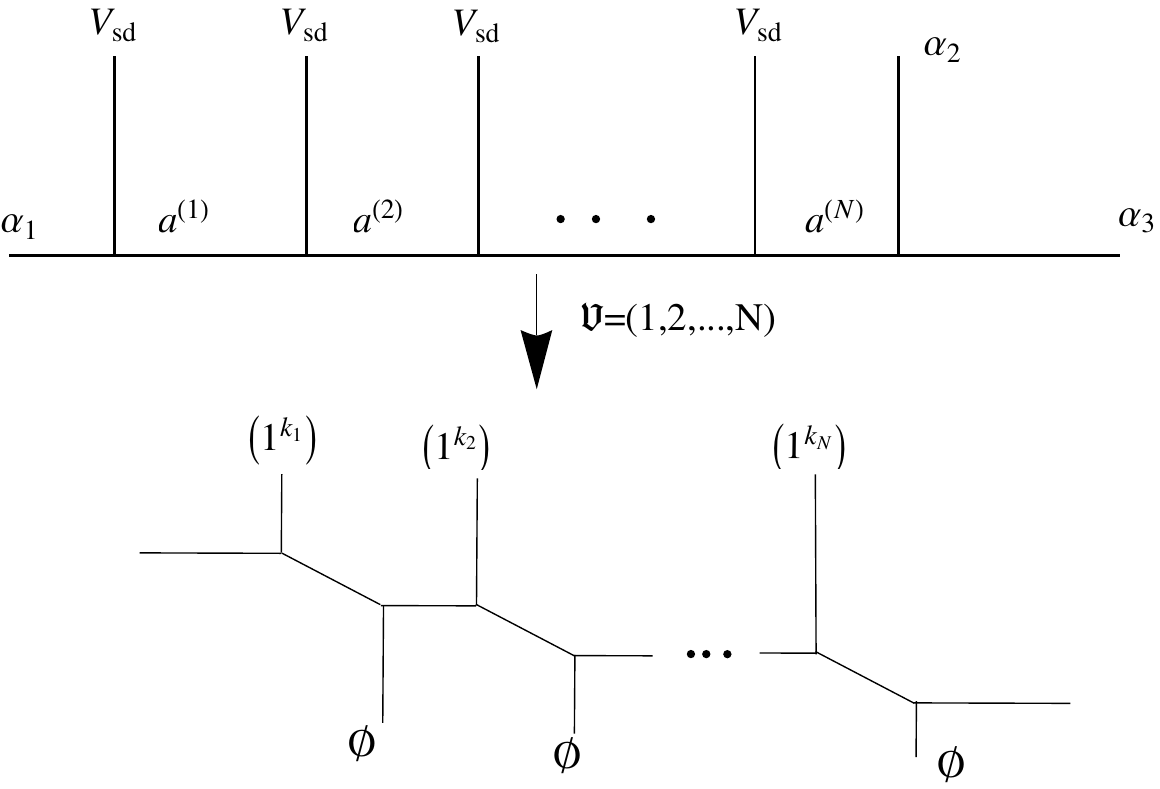}
\caption{relation between $SU(N)$ {}vortex and CFT}
\end{center}
\label{fig:sunvortex}
\end{figure}

\begin{figure}
\begin{center}
\includegraphics[ width=0.8\textwidth]{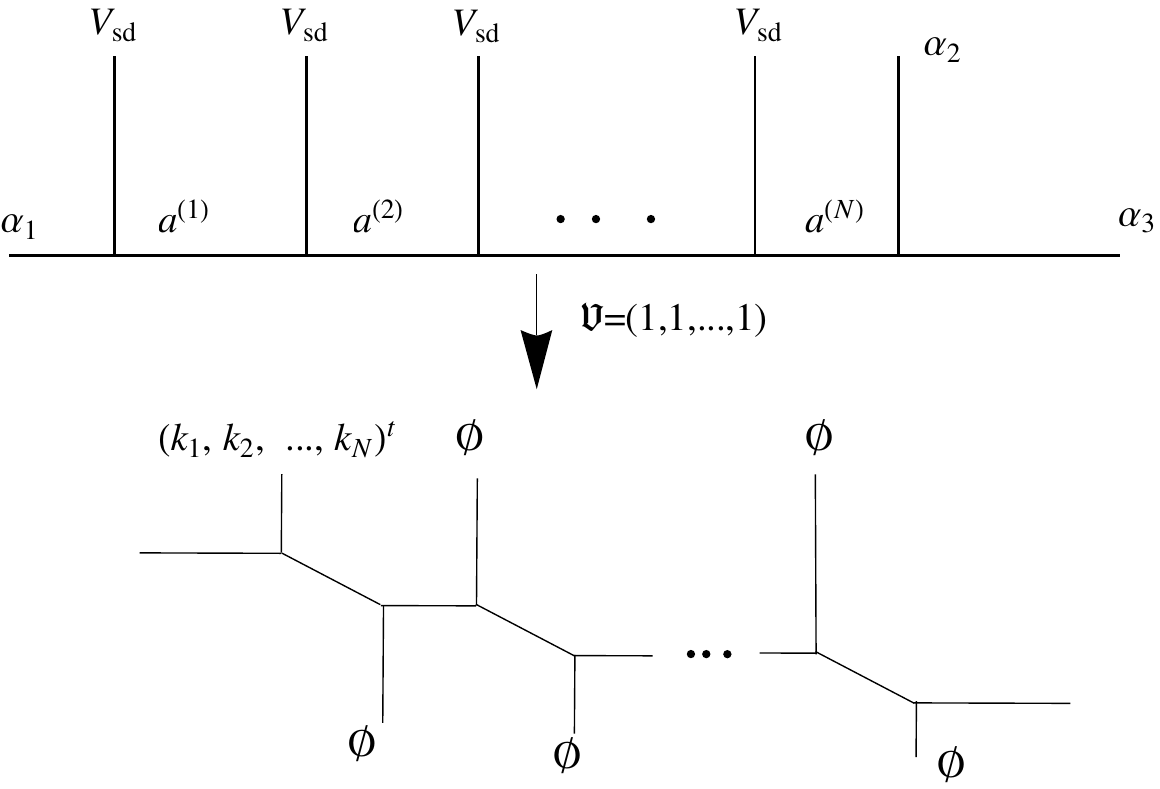}
\caption{relation between $SU(N)$  simple surface operators and CFT}
\end{center}
\label{fig:sunsurface}
\end{figure}

\begin{figure}
\begin{center}
\includegraphics[ width=0.8\textwidth]{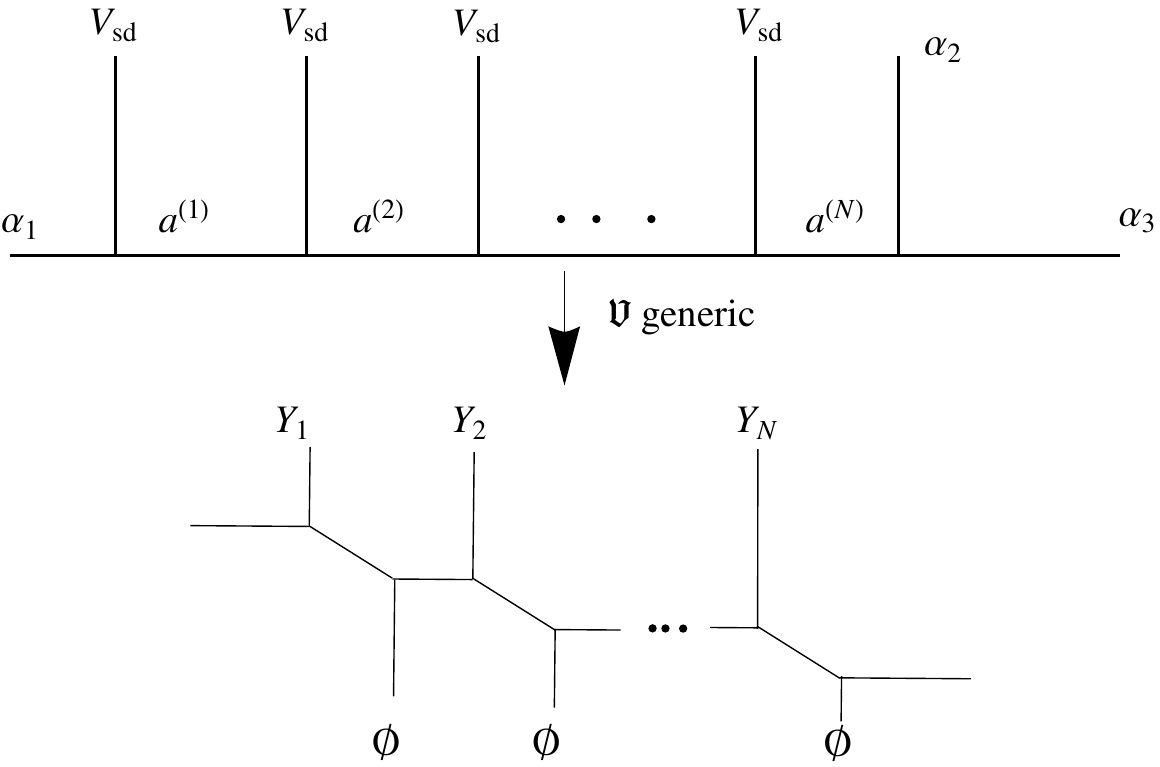}
\caption{relation between $SU(N)$  strip amplitudes and CFT}
\end{center}
\label{fig:sungeneral}
\end{figure}
Notice that the four dimensional limit of strip amplitudes correspond to conformal blocks with only two
independent external momenta, and one independent internal momentum. The number of degenerate states inserted in the conformal block corresponds to the total number of rows of the Young tableaux parametrizing the open string boundary conditions. This suggests that in order to have arbitrary boundary conditions one should consider conformal blocks
with an arbitrary number of degenerate field insertions. Since we know that the full instanton partition function
can be obtained by gluing two strip amplitudes with generic boundary conditions, this would provide a conformal field theory picture of this operation. From the CFT viewpoint, the infinite number of degenerate insertions could condense
in a line operator \cite{gaiottoline} which could be used to glue the two CFT amplitudes to obtain the full result.

\section{Conclusions}

In this paper we studied the relation between non-abelian vortex partition functions and Liouville/Toda conformal field theories,
by showing how to reproduce these partition functions from conformal blocks with degenerate field insertions.
Moreover, we performed a general analysis using geometric engineering for open topological strings and
found that there is a much richer structure in this correspondence which arises by identifying
the full vector space of degenerate conformal blocks with the four dimensional limit of open topological strings
amplitudes on a strip with general boundary conditions.
A natural generalisation of this approach would be to analyse the full refined topological string amplitudes from the CFT
viewpoint possibly along the lines of \cite{awa}.
Another interesting venue is the investigation of the correspondence with integrable systems and their quantization
\cite{hitchin,bmt2,cheng}
and in particular their
relevance in vortex counting and more in general for open topological string amplitudes on the strip.

As discussed in \cite{bucov,BTZ}, vortices partition functions arise in the classical limit of four dimensional gauge
theories with surface operator insertions. The analysis presented in this paper should then provide
the classical limit of multiple surface operator insertions. In particular the approach of quiver gauge theories
that we presented can be generalised to encompass also the four-dimensional instanton corrections.
This should be completed with a description of the moduli space of instantons with multiple defect insertions.

An analogous analysis could be performed for surface operators in gauge theories on ALE spaces
which have been recently related to para Liouville/Toda conformal CFTs \cite{bf,nt,bmt,bbb}.
In this case the relevant vortex moduli space should be obtained as a lagrangian submanifold
of the moduli space of instantons on ALE spaces.

\vspace{.5cm}
{\bf Acknowledgements}
We thank H.~Kanno, K.~Maruyoshi and S.~Pasquetti for interesting discussions and comments.
G.B. and Z.J. are partially supported by the INFN project TV12. A.T. is partially supported
by PRIN ``Geometria delle varietÂ´a algebriche e loro spazi di moduli'' and the INFN
project PI14.

\appendix
\section{Appendix}

\subsection{Instanton Partition Functions}

Let us consider
the instanton partition function of a linear quiver with N nodes. The corresponding brane construction
has $N+2$ sets of D4-branes and $N+1$ $NS5$ branes. We will focus on unrefined limit $\epsilon_1=-\epsilon_2$.
\be
Z_{\text{Quiver}}=Z_{\text{fund}}Z_{\text{antifund}}Z_N\prod _{i=1}^{N-1}Z_i Z_{i,i+1}
\ee
$Z_{\text{fund}}$ and $Z_{\text{antifund}}$ are the contributions from fundamental and antifundamental
hypermultiplets. $Z_i$ is the contribution from the $i$-th gauge factor, while $Z_{i,i+1}$ is the contribution
from the $i$-th bifundamental hyper.
These depend on the following parameters
\bea
a_i^{(N)} &=& \text{the i-th Coulomb branch parameter of the N-th gauge factor}.\nonumber\\
m_i &=& \text{the i-th mass of bifundamental hypermultiplet}\nonumber\\
\mu _i&=&\left\{
\begin{array}{cc}
 \text{masses of antifundamental hypermultiplets} & i\in [1,N] \nonumber\\
 \text{masses of fundamental hypermultiplets} & i\in [N+1,2N]
\end{array}
\right.\nonumber\\
Y^{(i)} &:& \text{the arrow of Young-tableaux on the i-th node. }\nonumber
\eea
More explicitly:
\bea
Z_{\text{antifund}}\left(a^{(1)},\mu,Y^{(1)}\right)&=&\prod _{f=1}^N\prod _{\alpha =1}^N\prod _{(i,j)\in Y_{\alpha }^{(1)}}\left(a_{\alpha }^{(1)}+\mu _f+\epsilon _2(j-i)\right)\\
Z_{\text{fund}}\left(a^{(N)},\mu,Y^{(N)}\right)&=&\prod _{f=1}^N\prod _{\alpha =1}^N\prod _{(i,j)\in Y_{\alpha }^{(N)}}\left(a_{\alpha }^{(N)}-\mu _{f+N}+\epsilon _2(j-i)\right)
\eea
The $L$-th bifundamental hypermultiplet contribution is:
\bea
Z_{L,L+1}&=&\prod _{\alpha =1}^N\prod _{\beta =1}^NZ_{L,L+1}^{(\alpha ,\beta )}\\
Z_{L,L+1}^{(\alpha ,\beta )}&=&\prod _{s\in Y_{\alpha }^{(L)}} \left(m_{\alpha ,\beta }^{(L)}+\epsilon _2\left(A_{Y_{\alpha }^{(L)}}(s)+L_{Y_{\beta }^{(L+1)}}(s)+1\right)\right)\nonumber\\
&&\prod _{t\in Y_{\beta }^{(L+1)}} \left(m_{\alpha ,\beta }^{(L)}-\epsilon _2\left(A_{Y_{\beta }^{(L+1)}}(t)+L_{Y_{\alpha }^{(L)}}(t)+1\right)\right)\nonumber\\
m_{\alpha ,\beta }^{(L)}&\text{:=}&a_{\alpha }^{(L)}-a_{\beta }^{(L+1)}-m_L
\eea
The $L$-th gauge factor contribution is:
\bea
Z_L &=&\prod _{\alpha =1}^N\prod _{\beta =1}^NZ_L^{(\alpha ,\beta )}\\
\left(Z_L^{(\alpha ,\beta )}\right)^{-1}&=&\prod _{s\in Y_{\alpha }^{(L)}} \left(a_{\alpha ,\beta }^{(L)}+\epsilon _2\left(A_{Y_{\alpha }^{(L)}}(s)+L_{Y_{\beta }^{(L)}}(s)+1\right)\right)\nonumber\\
&&\prod _{t\in Y_{\beta }^{(L)}} \left(a_{\alpha ,\beta }^{(L)}-\epsilon _2\left(A_{Y_{\beta }^{(L)}}(t)+L_{Y_{\alpha }^{(L)}}(t)+1\right)\right)
\eea
For a Young-tableau $Y$, one box $s$ has coordinates $(i,j)$, where $i$ counts the number of columns and $j$ counts the number of rows. Then the arm and leg of $s$ relative to another Young-tableau $W$,are defined as \(A_W(s)\text{:=}W_i-j
; L_W(s)\text{:=}W_j^t-i\). Where \(W^t\) is the dual partition of $W$. $|Y|\text{:=}\sum _iY_i$. We call
a partition of the form $(1^k)$ a row partition of length $k$, and a partition of the form $(k)$ a column partition.

\subsection{Degeneration from bifundamental masses}

Let us state our results and then prove them. The claim is that when \(m_{\alpha ,\beta }=0\) , \(W_{\beta }=Y_{\alpha }\) and when \(m_{\alpha ,\beta }=\epsilon _2\), \(W_{\beta }\) has one row more than that of \(Y_{\alpha}\). In this situation, if we suppose $Y_{\alpha}$ has $L$ rows with lengths
$k_1 \leq k_2 \leq \ldots\leq k_L$ and $W_{\beta}$ had $L+1$ rows with lengths $l_0 \leq l_1 \leq \ldots \leq l_L$, then for $1\leq i \leq L$ either
$k_i=l_{i-1}$ or $k_i=l_i+1$. Please refer to Fig.11 for a pictorial illustration.
\begin{figure}
\begin{center}
\includegraphics[ width=0.8\textwidth]{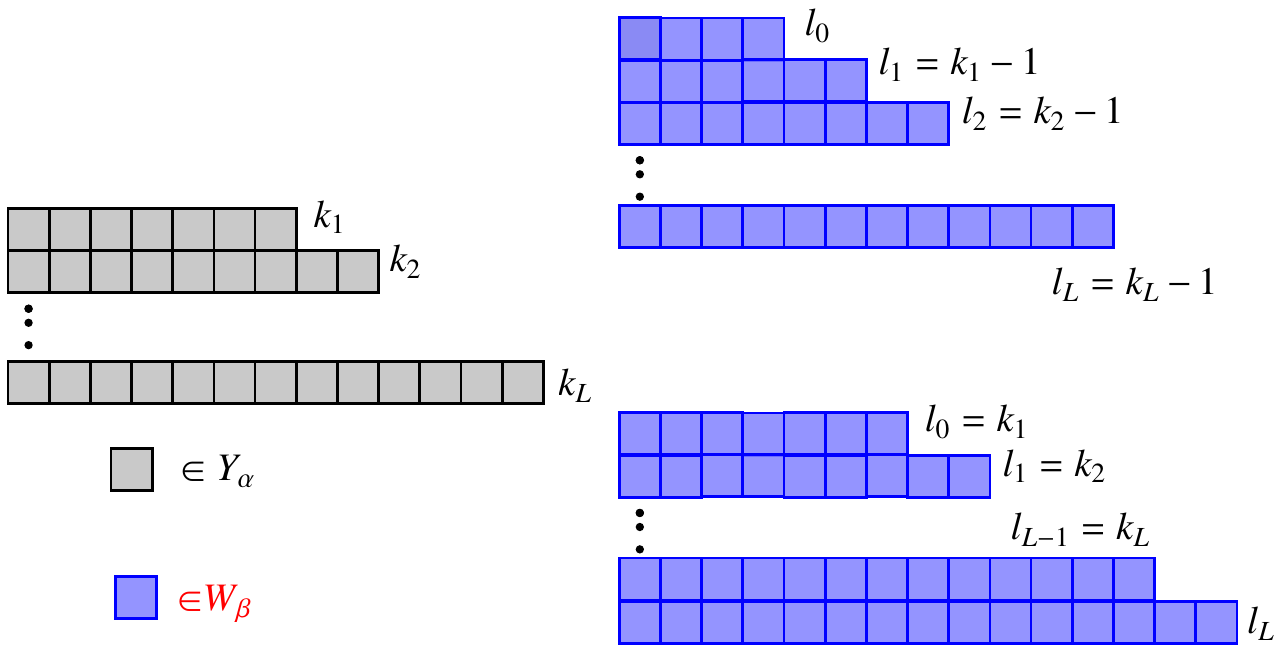}
\caption{Relation between $Y_{\alpha}$ and $W_{\beta}$ when $m_{\alpha,\beta}=\epsilon_2$}
\end{center}
\label{fig:young}Then consider contribution from s.
\end{figure}
Let's start from the simpler case \(m_{\alpha ,\beta }=0\).
\be
Z_{\text{bifund}}^{(\alpha ,\beta )}=\prod _{s\in Y_{\alpha }} \left(\epsilon _2\left(A_{Y_{\alpha }}(s)+L_{W_{\beta }}(s)+1\right)\right)\prod
_{t\in W_{\beta }} \left(-\epsilon _2\left(A_{W_{\beta }}(t)+L_{Y_{\alpha }}(t)+1\right)\right)\nonumber
\ee
Let's suppose $ Y_{\alpha }^t=\left(k_1,k_2,\text{...},k_L\right); W_{\beta }^t=\left(l_1,l_2,\text{...},l_M\right)$.
We will proceed in our proof by induction from the top row to the bottom.

If \(M>L\), then the result is non vanishing only if
\(L_{Y_{\alpha }}(t)+1=L_{\emptyset }(t)+1=-i+1\neq 0\).
The same argument applies for \(L\leq M\), so that we stay with \(M=L\).

The first induction step is
when \(t\) is on the top row of \(W_{\beta }\), so that \(A_{W_{\beta }}(t)=0\), and \(L_{Y_{\alpha }\text{}}(t)+1=1+k_1-i_t\)
\be
 \left\{
\begin{array}{c}
 L_{Y_{\alpha }\text{}}(t)+1=1+k_1-i_t\neq 0 \\
 i_t\in \left[1,l_1\right]
\end{array}
\right. \Longrightarrow k_1\geq l_1 \nonumber
\ee
Similarly for the contribution from \(s\in Y_{\alpha }\), we get \(l_1\geq k_1\), implying \({k_1=l_1}\).
Suppose now \(k_i=l_i\) when \(i\leq p-1\) and  let's prove that \(k_p=l_p\).

1. \(i_t\in \left[1,l_1\right]\), \(A_{W_{\beta }}(t)=p-1\)
\be
\left\{
\begin{array}{c}
 L_{Y_{\alpha }\text{}}(t)+p=p+k_p-i_t\neq 0 \\
 i_t\in \left[1,l_1\right]
\end{array}
\right. \Longrightarrow k_p\geq l_1-(p-1)\nonumber
\ee

2. when \(i_t\in \left[l_1+1,l_2\right]\) £¬\(A_{W_{\beta }}(t)=p-2\)
\be
\left\{
\begin{array}{c}
 L_{Y_{\alpha }\text{}}(t)+1+p-2=p-1+k_p-i_t\neq 0 \\
 i_t\in \left[1+l_1,l_2\right]
\end{array}
\right. \Longrightarrow k_p\leq l_1+1-p\text{  }\text{or}\text{  }k_p\geq l_2-p+2 \nonumber
\ee
Since \(k_1=l_1\), then \(k_p\neq l_1+1-p\) and \(k_p\geq l_2-p+2\).
By iterating this procedure we find \(k_p\geq l_p\), and symmetrically \(l_p\geq k_p\), namely \(l_p=k_p\).
This ends the proof of the first statement.

Now let us concentrate on \(m_{\alpha ,\beta }=\epsilon _2\)
\be
Z_{\text{bifund}}^{(\alpha ,\beta )}=\prod _{s\in Y_{\alpha }} \left(\epsilon _2\left(A_{Y_{\alpha }}(s)+L_{W_{\beta }}(s)+2\right)\right)\prod
_{t\in W_{\beta }} \left(-\epsilon _2\left(A_{W_{\beta }}(t)+L_{Y_{\alpha }}(t)\right)\right)\nonumber
\ee
It is easy to show that \(W_{\beta }\) can have at most one row more than \(Y_{\alpha }\).
Suppose that $ Y_{\alpha }^t=\left(k_1,k_2,\text{...},k_L\right); W_{\beta }^t=\left(l_0,l_1,l_2,\text{...},l_M\right)$ and apply
induction again from top to bottom.


When $t$ is on the top row of \(W_{\beta }\) there is no constraint for the length \(l_0\).


When \(t\) is on the next to top row of \(W_{\beta }\) then:

1. for \(i_t\in \left[1,l_0\right]\), in this case \(A_{W_{\beta }}(t)=1\)
\be
\left\{
\begin{array}{c}
 L_{Y_{\alpha }\text{}}(t)+1=1+k_1-i_t\neq 0 \\
 i_t\in \left[1,l_0\right]
\end{array}
\right.\Longrightarrow k_1\geq l_0\nonumber
\ee

2. for \(i_t\in \left[l_0+1,l_1\right]\), in this case \(A_{W_{\beta }}(t)=0\)
\be
\left\{
\begin{array}{c}
 L_{Y_{\alpha }\text{}}(t)=k_1-i_t\neq 0 \\
 i_t\in \left[1+l_0,l_1\right]
\end{array}
\right.
\Longrightarrow k_1\leq l_0\text{  }\text{or} k_1\geq l_1 +1\nonumber
\ee
so we have { \(k_1=l_0\)}{  or }{ \(k_1\geq l_1+1\)}.

Let us consider now the contribution from $Y_\alpha$.
When $s$ is on the top row of \(Y_{\alpha }\), { }\(A_{Y_{\alpha }}(s)=0\)
and we get
\be
\left\{
\begin{array}{c}
 L_{W_{\beta }}(s)+2=l_1-i_s+2\neq 0 \\
 i_s\in \left[1,k_1\right]
\end{array}
\right.\Longrightarrow {l_1}{\geq }{k_1-1}\nonumber
\ee
so \(k_1=l_0\) or \(k_1\geq l_1+1\) $\cap $ \(l_1\geq k_1-1\) $\Longrightarrow ${  }{ \(k_1=l_0\)}{  or }{ \(k_1=l_1+1\)}.
Now suppose that for \(i\leq p-1\), we have \(l_i=k_i-1\) or \(l_{i-1}=k_i\). Then

1. for \(i_t\in \left[1,l_0\right]\), \(A_{W_{\beta }}(t)=p\) and
\be
\left\{
\begin{array}{c}
 L_{Y_{\alpha }\text{}}(t)+p=p+k_p-i_t\neq 0 \\
 i_t\in \left[1,l_0\right]
\end{array}
\right.\Longrightarrow k_p\geq l_0-p+1\nonumber
\ee

2. for \(i_t\in \left[l_0+1,l_1\right]\), \(A_{W_{\beta }}(t)=p-1\)
\be
\left\{
\begin{array}{c}
 L_{Y_{\alpha }\text{}}(t)+p-1=p-1+k_p-i_t\neq 0 \\
 i_t\in \left[1+l_0,l_1\right]
\end{array}
\right.\Longrightarrow k_p\leq l_0+1-p\text{   }\text{or}\, k_p\geq l_1+2-p\nonumber
\ee
so we have \(k_p=l_0+1-p\) or \(k_p\geq l_1+2-p\)
By iterating this procedure we get :
\(k_p=l_0+1-p\text{  }\text{or}\, k_p=l_1+2-p,\text{...},\text{or}\, k_p=l _{p-1}\text{      }\text{or} \, k_p\geq l_p+1\). From the induction assumption
we have \(k_p\geq l_p+1\)  or  \(k_p=l_{p-1}\).

Let us now consider the contribution from $s\in Y_\alpha$.

1.  for $i_s \in \left[1,k_1\right]$, \(A_{Y_{\alpha }}(s)=p-1\)
\be
\left\{
\begin{array}{c}
 L_{W_{\beta }}(s)+1+p=l_p-i_s+1+p\neq 0 \\
 i_s\in \left[1,k_1\right]
\end{array}
\right.\Longrightarrow
l_p\geq k_1-p\nonumber
\ee

2. for $i_s \in \left[k_1+1,k_2\right]$, \(A_{Y_{\alpha }}(s)=p-2\)
\be
\left\{
\begin{array}{c}
 L_{W_{\beta }}(s)+p=l_p-i_s+p\neq 0 \\
 i_s\in \left[k_1+1,k_2\right]
\end{array}
\right.\Longrightarrow
l_p\leq k_1-p\text{  }\text{or} l_p\geq k_2-p+1\nonumber
\ee
so we find \(l_p=k_1-p\) or \(l_p\geq k_2-p+1\).
By iterating the procedure
we find \(l_p=k_1-p\text{  }\text{or}\, l_p=k_2-p+1,\text{...}, \text{or}\, l_p=k_{p-1}-2\text{  }\text{or}\, l_p\geq k_p-1\).
From the induction assumption it follows that \(l_p\geq k_p-1\).

Finally, combining the results from $W_\beta$ and $Y_\alpha$, we have :
\(k_p=l_p+1\) or  \(k_p=l_{p-1}\), which is what we wanted to prove.

\subsection{Factorization formulae}

When \(Y_L^t=\left(l_1+1,l_2+1,\text{...},1+l_L\right)\) { }\(Y_{L+1}^t=\left(l_0,l_1,l_2,\text{...},l_L\right)\) { } (\(l_i\leq l_{i+1}\))
\bea
\left(\epsilon _2\right)_{Y_L,Y_{L+1}}&=&\prod _{(i,j)\in Y_L} \left(\epsilon _2+\epsilon _2\left(A_{Y_L}(i,j)+L_{Y_{L+1}}(i,j)+1\right)\right)\nonumber\\
&&\prod
_{(a,b)\in Y_{L+1}} \left(\epsilon _2-\epsilon _2\left(A_{Y_{L+1}}(a,b)+L_{Y_L}(a,b)+1\right)\right)\nonumber\\
&=&\prod _{(i,j)\in Y_L} \left(\epsilon _2+\epsilon _2\left(A_{Y_L}(i,j)+L_{Y_L}(i,j)\right)\right)\nonumber\\
&&\prod _{(a,b)\in Y_{L+1}\cap Y_L} \left(\epsilon
_2-\epsilon _2\left(A_{Y_{L+1}}(a,b)+L_{Y_{L+1}}(a,b)+2\right)\right)\nonumber\\
&&\prod _{(a,b)\in Y_{L+1}\backslash Y_L} \left(\epsilon _2-\epsilon _2\left(A_{Y_{L+1}}(a,b)+L_{Y_L}(a,b)+1\right)\right)\nonumber\\
&=& H_{Y_L}\prod _{(a,b)\in Y_{L+1}\cap Y_L} -h_{Y_{L+1}}(a,b)\prod _{(a,b)\in Y_{L+1}\backslash Y_L} h_{Y_{L+1}}(a,b)\nonumber\\
&=&(-1)^{\left|Y_L\right|-L}H_{Y_L}H_{Y_{L+1}} \nonumber
\eea
Similarly ,when \(Y_L^t=\left(l_1,l_2,\text{...},l_L\right)\) and \(Y_{L+1}^t=\left(l_1,l_2,\text{...},l_{L+1}\right)\) { } { }\(\left(l_i\leq l_{i+1}\right)\)
\bea
\left(\epsilon _2\right)_{Y_L,Y_{L+1}}&=&\prod _{(i,j)\in Y_L} \left(\epsilon _2\left(\left(A_{Y_L}(i,j)+1\right)+L_{Y_{L+1}}(i,j)+1\right)\right)\nonumber\\
&&\prod
_{(a,b)\in Y_{L+1}} \left(-\epsilon _2\left(\left(A_{Y_{L+1}}(a,b)-1\right)+L_{Y_L}(a,b)+1\right)\right)\nonumber\\
&=&\prod _{(i,j)\in Y_L} h_{Y_{L+1}}(i,j)\prod _{(a,b)\in Y_L} (-1)^{\left|Y_L\right|}h_{Y_L}(a,b)\nonumber\\
&&\prod _{(a,b)\in Y_{L+1}\backslash Y_L} \left(-\epsilon
_2\left(\left(A_{Y_{L+1}}(a,b)-1\right)+L_{Y_L}(a,b)+1\right)\right)\nonumber\\
&=&(-1)^{\left|Y_L\right|} H_{Y_L}H_{Y_{L+1}}\nonumber
\eea

\end{document}